# Proposed Quality Evaluation Framework to Incorporate Quality Aspects in Web Warehouse Creation

Umm-e-Mariya Shah, Maqbool Uddin Shaikh, Azra Shamim, Yasir Mehmood

**Abstract**— Web Warehouse is a read only repository maintained on the web to effectively handle the relevant data. Web warehouse is a system comprised of various subsystems and process. It supports the organizations in decision making. Quality of data store in web warehouse can affect the quality of decision made. For a valuable decision making it is required to consider the quality aspects in designing and modelling of a web warehouse. Thus data quality is one of the most important issues of the web warehousing system. Quality must be incorporated at different stages of the web warehousing system development. It is necessary to enhance existing data warehousing system to increase the data quality. It results in the storage of high quality data in the repository and efficient decision making. In this paper a Quality Evaluation Framework is proposed keeping in view the quality dimensions associated with different phases of a web warehouse. Further more, the proposed framework is validated empirically with the help of quantitative analysis.

**Index Terms**—Data Warehouse, Web Warehouse, Quality Assessment, Quality Evaluation Framework, WWW

——————————— ◆ ———————————

## 1 INTRODUCTION

Data is tremendously growing on the web. The operational and archived data exists on the web in a massive amount. World Wide Web has grown to be a universal source and is globally used by almost all individuals and business organizations for information sharing and exchange. The web space is diverse, distributed, heterogeneous and semi-structured in nature. Relevant information retrieval is difficult from the web space as there is no universal configuration and organization of the web data [1]. At present information retrieval mechanism is comprised of two tactics i.e. browsers and search engines. There are certain shortcomings of this information retrieval phenomenon. One of the shortcomings is that web browsers usually make use of the hyperlinks among web pages. Many search engines have limited knowledge or they are unaware of the use of the link information. As a result search engines are not able to support such queries or fail to return link information [2]. Another inadequacy is that there is no mechanism for coupling the relevant web documents returned by the search engines. User may manually visit and download the documents [2].

Most popular web servers become overloaded by the stream of the client's requests. Also Web servers can not keep track of the diverse behaviour of the client's requests and does not offer the services of web personalization. Therefore an intermediate storage area between the web servers and the clients proves to be a valuable resource [3]. The intermediate repository may not only serve as the storage area but also keeps track of the client's activities and helps in the web personalization [4]. Web warehousing can be used to resolve the problem.

In this research work the author has proposed a quality evaluation framework for a web warehousing system. The framework presents certain quality dimensions with respect to different phases of a web warehouse. These attributes will help to measure the features of the system and ensure quality of end results.

The rest of paper is organized into different sections. Section 2 provides the Literature Review regarding Data Warehouse and Web Warehouse. Section 3 is comprised of the discussion with respect to quality and Web Warehouse. The proposed quality evaluation framework is shown in section 4. The validation of the suggested framework is given in section 5. Section 6 present concluding marks.

## 2 LITERATURE REVIEW

### 2.1 Data Warehouse

Data Warehouse is a database maintained for business decision support. It is independent of the organization's operational databases. According to Hoffer et. al. "A data warehouse (DWH) is an 'informational database' that is

————————————————

- Umm-e-Mariya Shah is a student at COMSATS Institute of Information Technology, Islamabad. Pakistan
- Maqbool uddin shaikh is a professor at COMSATS Institute of Information Technology, Islamabad. Pakistan
- *Azra Shamim* is working as a research associate in COMSATS Institute of Information Technology, Islamabad, Pakistan
- Yasir Mehmood is Senior Software Engineer at Buraq Technologies, Wah Cantt. Pakistan

maintained separately from an organization's operational database" [5]. "*A collection of corporate information, derived directly from operational systems and some external data sources. Its specific purpose is to support business decisions, not business operations*" [6].

According to Inmon, "*A Data Warehouse is a subject-oriented, integrated, time-variant, non volatile collection of data in support of management decisions*" [7]. In data warehouse, data from heterogeneous sources is extracted, transformed and loaded into a separate storage place. Thus providing a platform for direct querying and analysis, from historical perspective. Basic architecture of a data warehouse is discussed in [8]. In this architecture heterogeneous data from dissimilar sources is put together into a warehouse.

## 2.2 Web Warehouse

A Web warehouse is composed of data warehousing technology embedded with the web technology. According to Mattison, "It is an approach to the building of computer systems which has as its primary functions the identification, cataloguing, retrieval, (possibly) storage, and analysis of information (in the form of data, text, graphics, images, sounds, videos, and other multimedia objects) through the use of Web technology, to help individuals find the information they are looking for and analyse it effectively" [9]. This definition shows that a web warehouse is an architecture that provides some tools and processes to deal with heterogeneous data like text, numeric, graphical data etc. The main sources of a web warehouse are the web sites. Web warehousing only deals with managing and organizing the stored data. Information sharing and intelligent caching are the key factors of web warehousing system. Web warehouse provides comprehensive and up-to-date information about different domains. It offers powerful support to the users or communities for sharing and exchange of information. It helps the business analysts in an efficient knowledge discovery.

## 2.3 Contributors of Web Warehouse

As Data Warehousing technology blends into web warehousing so it contributes two major characteristics towards it [10] as shown in figure 1.

- **Data Warehouse as a Contributor to Web Warehouse**

One of the contributions is the managerial and organizing principles. Web warehouse manages the stored information in the same way as done by data warehouse. Another contribution is that similar database applications run in web warehouse systems as in data warehouse systems e.g. OLAP operations, data mining practices, and statistical analysis techniques.

- **Web Technology as a Contributor to Web Warehouse**

Web technology is another contributor to web warehouse [10]. Web technology introduces new approaches to convey information to the users. With respect to web warehousing it gives an ease to business analysts to search, access, analyse and manipulate the non-data-based data in a similar way as they process data-based data.

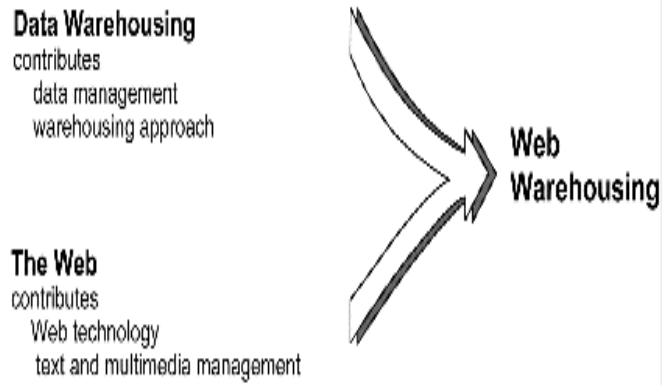

Fig. 1 Main Contributors of Web Warehousing [10]

## 2.4 Features of a Web Warehouse

Web Warehousing supports some features that help business analysts and other users to analyse well and make effective decisions. Following are some of the features of web warehouse.

- **Web Mining**

The web contains valuable information so it serves to be fruitful for mining useful knowledge. Digging out the relevant information from the huge contents of web is termed as Web Mining. [11]

- **Web Analysis**

Web analysis means to examine data over the web. Web Servers maintain web logs to store the information of access request. For each browsing session information is recorded in log files.

- **Archiving Web Data**

Web warehouse maintains archives for the web data. It is an important feature of web warehousing. So web archives help in making valuable business decisions and provide a platform for heterogeneous data gathered from various websites and stored in a unified format.

## 2.5 Architectures of a Web Warehouse

Various architectures are proposed by different researchers in the literature. In [12] an overview of the basic and main functionality of a web warehouse was described. It shows that there are three main parts of the web warehouse architecture: input, storage and output. The input part is responsible for acquiring useful information which is filtered/ clustered and stored in the storage area. The storage area is responsible for the data manipulation and organization. The output area interlinks with the storage area. It uses the stored information and presents it to the user via proper interface comprising of the analytical tools.

Saif et. al. discussed basic architecture of web warehouse in [13]. In [14] Lean Yu et. al. presented a general frame work of a web warehouse for the decision support system. The proposed architecture illustrates that different elements are involved in the creation of a web warehouse. Layered Architecture of a web warehouse is proposed by Yan Zhu and Alejandro Buchmann in [15] . Yan Zhang et. al [16] put forward a new architecture of web warehouse. In this architecture the acquirement process of information is presented in a different way involving monitors/wrappers, integrator and view managers.

## 3 QUALITY AND WEB WAREHOUSE

Web technology is one of the contributors towards a web warehousing system besides data warehousing technology [10]. Web technology introduces new approaches to convey the information to the users. With respect to the web warehousing system it gives an ease to the business analysts to search, access, analyse and manipulate the non-data-based data in a similar way as they process the data-based data.

The involvement of the Web with a warehouse increases the challenges of quality. This is because of the semi-structured nature of the web data, lack of control over the data sources and frequently changing data on these sources. The nature of the web itself creates problems related to quality. Some of these problems affecting quality can be design problems or out of date/un–updated information. These issues make it difficult for the users to easily access the information or make use of the published information which is inconsistent and contains obsolete links, respectively.

As data is the key factor for powerful and accurate decision making so it must be of high quality. Data quality can be defined as "fitness for the use and the fraction of performance over expectancy" [17]. Data semantics are different for each distinct user and these users are the best evaluators for the quality of data. End users require coherency, freshness, accuracy, accessibility, availability and performance in terms of the quality data for a warehouse. The timeliness and the ease of querying a web warehouse are the attractive features for a decision maker. On the other hand an administrator of the web warehouse would like to see the quality of metadata, data currency and efficient error reporting etc. Thus quality has diversified features and goals according to each user.

Complex measurements, predictions, and design techniques are required for the quality evaluation. Objective quality factors are computed and compared with each user's expectation in order to predict the data quality. Integration of different web sources and the automation of the conceptual design for multiple web sources are also considered. For each user this quality assessment is a very challenging task in order to satisfy each user with different and sometimes opposing requirements. Prioritization of these requirements is also challenging as large number of users exists for a warehouse.

## 4 PROPOSED QUALITY EVALUATION FRAMEWORK FOR A WEB WAREHOUSING SYSTEM

Quality is one of the important factors for the success and survival of any system. Web Warehouse is a system that helps in the analysis and decision making in any organisation. It is required that the implementation of the Web Warehousing System must be followed with proper evaluation/validation of each phase against certain attributes that measure the quality of the system. It helps to get better end results and thus leads to the effective decisions. From [18], [19],[20] and [21] a framework is suggested below to assess the quality of a web warehousing system. Proposed quality evaluation framework is shown in figure 2.

The description of the framework on the basis of the categories and dimensions of the quality factors is given in the following table. It also shows the relevancy of each category to the phases and sub-phases of a Web Warehouse. The proposed attributes help to evaluate each phase of a Web Warehouse.

The description of the proposed quality attributes are as follow:

- Accessibility

This attribute is defined in terms of privileges it measures that whether proper rights are defined for the users of a web warehousing system or not.

- Interpretability

It measures how clearly the system's plan and all scenarios are reflected. For instance in case of query processing it makes certain the clarity of meaning, structure and language rules. The design must also reflect the ease with which the origin/source of data can be traced and a user understands the concept of data for analysis purpose.

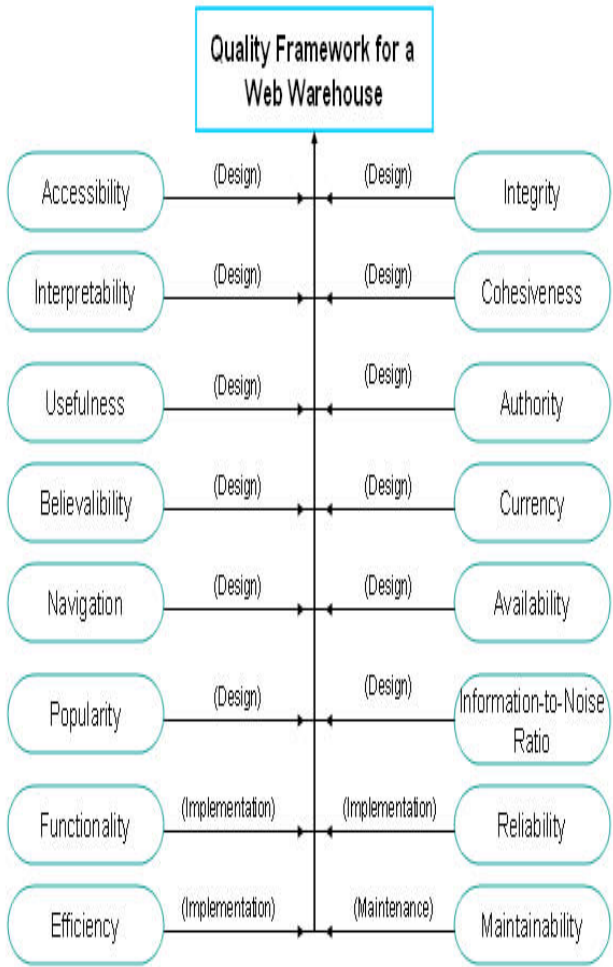

Fig. 2 Quality Evaluation Framework for a Web Warehouse

TABLE I
PROPOSED FRAMEWORK TO EVALUATE THE PHASES OF A WEB WAREHOUSE

| Phases | Sub-Phases | Categories | Dimensions |
|---|---|---|---|
| Creation | Design | Accessibility | Privileges |
| | | Interpretability | Semantics, Syntax, Origin |
| | | Usefulness | Clarity, Understandability, Explicitness, Relevancy, Flexibility, Usage, Timeliness, Value-added |
| | | Believability | Accuracy, Completeness, Consistency, Credibility |
| | | Navigation | Intuitive Design |
| | | Efficiency | Performance, Time Behaviour, Optimization |
| | | Authority | ---- |
| | | Currency | ---- |
| | | Availability | ---- |
| | | Information-to-Noise Ratio | ---- |
| | | Popularity | ---- |
| | | Cohesiveness | ---- |
| | | Integrity | Redundancy |
| | Implementation | Reliability | Fault tolerance, Recoverability, Availability |
| | | Functionality | Suitability, Accuracy, Interoperability, Compliance, Security, Traceability |
| | | Efficiency | Time behaviour, Resource behaviour |
| Maintenance | | Maintainability | Analyzability, Changeability, Stability, Testability, Manageability, Reusability |

- Usefulness

It evaluates if a system's design possesses a set of comprehensive and reasonable rules relevant to the requirements of the system and flexible enough to made modifications. It also measures the ease with which data can be used for analysis, applicable for a certain situation, up-to-date and assist in decision making.

- Believability

The quality constructs believability measures whether the objective and subjective components of the web warehouse are properly and comprehensively defined. In terms of data quality assessment it also assesses that data must be correct, comprehensive, precise and stable.

- Navigation

It measures the degree of reflection of an insightful meaning of the design.

- Efficiency

In case of Query Processing it measures that if the web warehousing system timely responds to an assigned task.

- Authority

It determines that whether the web warehousing system is able to effectively measure the prestige of a selected source during the source selection process.

- Currency

This attribute assesses the source selection process. It measures whether the selected source has up-to-date contents or not.

- Availability

During the source selection process it measures the number of broken links.

- Information-to-Noise Ratio

It measures the proportion of the useful information in a selected source.

- Popularity

This quality factor determines the number of citations by other web pages and helps to evaluate the selected source.

- Cohesiveness

It computes the relevancy of the major topics in a web page, while source evaluation phenomenon.

- Integrity

It measures that the system's design should be strong enough to ascertain that data will not repeat.

- Reliability

This attribute determines that the system could not crash under unexpected condition and can recover the losses.

- Functionality

It measures that the system is secure, up to the standards, complete according to the requirements and inter-operate in a diverse system environment.

- Efficiency

Efficiency measures that the system has enough resources to timely respond to the intended task.

- Maintainability

It evaluates that how much the web warehousing system is securely established, under control, flexible enough to add new contents and reuse the existing contents.

## 5 VALIDATION OF THE PROPOSED FRAMEWORK

The worth of a proposed framework can be validated empirically with the help of an approach of quantitative analysis. It is termed as "validity analysis". In this approach a framework is validated on the basis of three quality factors: Sufficiency/Completeness, Necessity/Parsimony and Independence/Orthogonal [22], [23].

Sufficiency/Completeness measures that whether the proposed set of quality attributes are adequate to evaluate the overall quality of a web warehouse or there is any missing attribute. Similarly the factor Necessity/Parsimony ensures that all the defined attributes are obligatory and effectively contribute towards the complete set of system's quality evaluation framework. Independence/Orthogonal guarantees that all attributes are independent and there should be no overlapping between them. If a framework comes up to the mark according to these three aspects then it will be a valid framework.

For the validation phenomenon the quality constructs from 5 generally recognized quality frameworks are considered as the empirical observations [21]. These quality factors are then mapped with the proposed factors. The mapping procedure is shown in the figure 3.

The figure 3 shows that the proposed set of attributes is not sufficient because three quality factors (highlighted in green in the above figure) from the empirical observations set did not map to any of the proposed factors. These attributes must be included in the defined quality framework for a web warehouse. The validation also shows that all the suggested factors are necessary for the evaluation of the overall system's quality, as all the proposed quality constructs map to the constructs found in the literature. Empirical validation also makes it clear that the recommended attributes are not independent as some of the quality factors from the set of empirical observations map to more than one construct in the suggested set of factors thus showing overlapping between the attributes.

The overlapping attributes are highlighted in red in the above figure. The overlapping can be justified if there is a clear distinction between the definitions of the factors. The overlapping attributes can be shown more clearly in figure 4.

Each overlying quality dimension has its own distinct definition. For instance the quality attribute "Contextual", maps with the three constructs "Usefulness", "Believability" and "Integrity". Usefulness defines relevancy in terms of system's design. It measures that the design possesses such rules that must be relevant to the requirements of the organization's warehousing system. Timeliness and Value-added represents the data quality. Timeliness measures that the data is appropriate for analysis and Value-added shows that data is up-to-date. Concerning the category "believability" its dimension "completeness" shows that the objective and subjective components of the web warehousing system are fully defined. "Integrity" measures that the redundancy if data in a web warehousing system.

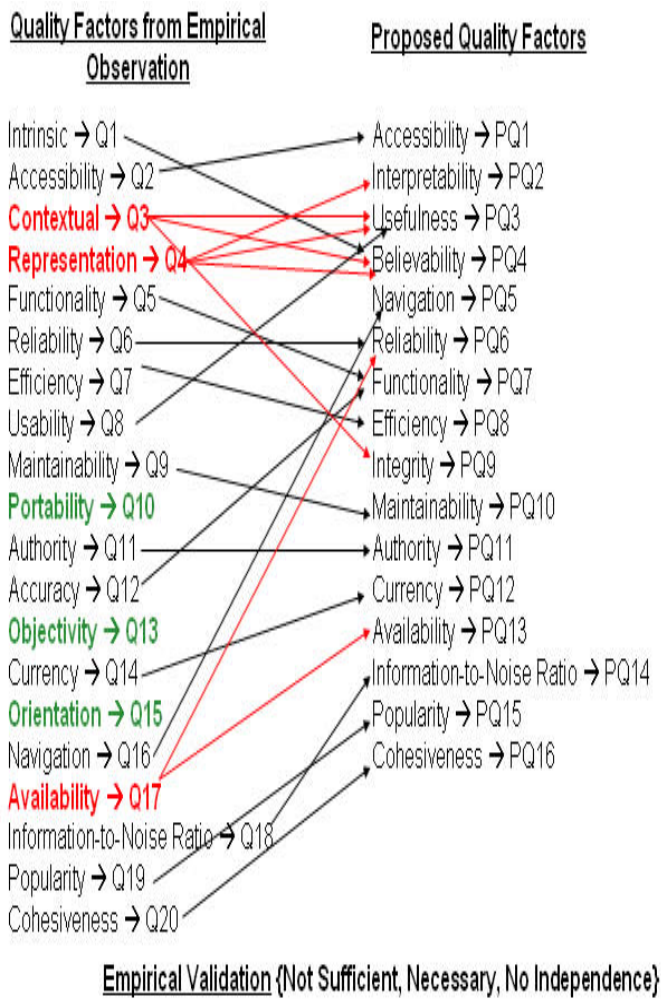

Fig. 2 Empirical Validation of the Proposed Framework

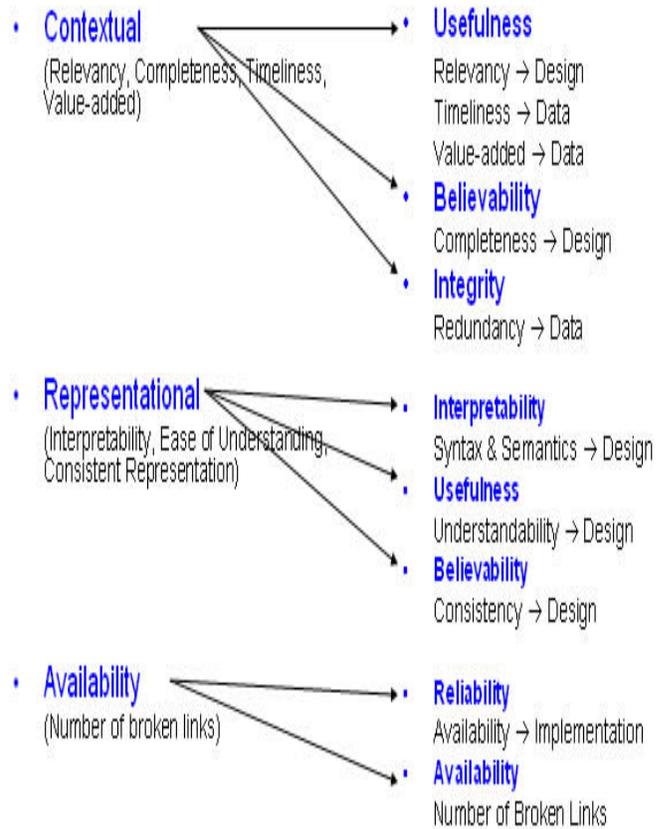

Fig. 3 The Overlapping Quality Attributes in the Proposed Framework

Similarly the above figure shows that the attribute "Representational" maps to Interpretability, Usefulness and Believability of the proposed framework. The three attributes determines the quality of design.

Where interpretability shows how clearly the syntax and semantics of the system's design are defined, usefulness gives the comprehensible and logical view of design and believability defines the steadiness and reliability of the system.

The attribute "Availability" from the empirical observations overlaps with "Reliability" and "Availability" of the suggested quality constructs. Reliability is an implementation issue and ensures the availability of the system in case of failures or any severe/unexpected conditions. Whereas the second factor Availability represents the number of broken links which are measured at the time of source evaluation.

On the basis of the above validation phenomenon the revised set of quality factors for a web warehouse system are shown in figure 5. The new additions are highlighted in red thus justifying the sufficiency of the framework.

This layer assesses the origin of the data that is selected for data extraction. Assessment is on the basis of some quality dimensions i.e. source currency, relevancy, availability, information-to-noise ratio, authority, popularity and cohesiveness. It measures the number of broken links on a web page, proportion of the useful information, prestige of the data source, relevancy of the major topics in a web page and the number of citations by other web pages.

The evaluation ensures that there exist up-to-date contents in the selected source that are compatible to the user's query. The source evaluation layer then filters out the data belonging to those sources only that satisfies the evaluation criteria.

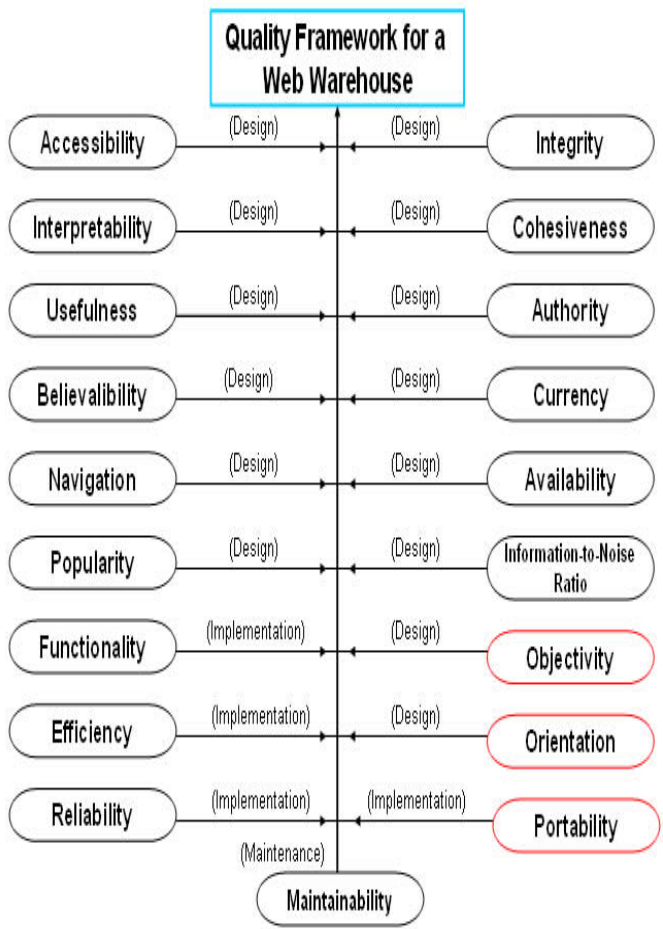

Fig. 4  Improved Quality Framework for a Web Warehouse

1. Monitor

This component is connected to the underlying information sources. Each data source has its monitor. It polls the web information sources periodically to detect any changes arise in them. Polling is done by comparing the snapshots and obtaining the base data changes. It collects the changes and notifies the modifications to the integrator component.

2. Wrapper

This component deals with the data extraction. It accepts the query from query processor and mines results from the underlying information sources. The result is then transformed into a specified format of a web warehousing system.

3. Integrator

This component maintains consistency between the web warehousing system and the underlying information resources. Any updated information is sent to the integrator by the monitor. Integrator integrates the information and sends the modifications to the respective view manager.

## 6 CONCLUSION

Web warehouseing systems support organizations in strategic decision making. Qualilty of data store in web warehouse affect quality of decion made by top management.  So quality assessment is most important at various stages of the web warehouse to get effective web warehousing system. In this paper a quality evaluation framework is proposed and discussed to check the quality dimensions of web warehouse. The worth of the proposed frame work is check by empirical validation.

**Umm-e-Mariya Shah** is a student of MS CS in COMSATS Institute of Information Technology. In addition she is an IT Consultant in Sustainalbe Development Policy Institute, Islamabad. Pakistan. Also she is a visiting lecture in SKANS School of Accountancy, Rawalpindi, Pakistan.

**Maqbool Uddin Shaikh** is a foreign faculty member in computer science department of COMSATS Institute of Information Technology, Islamabad, Pakistan. He received his PhD degree from Liver Pool University.

**Azra Shamim** is working as a research associate in COMSATS Institute of Information Technology, Islamabad, Pakistan. She received her MS CS degree from COMSATS Institute of Information Technology, Islamabad, Pakistan.

**Yasir Mehmood** is working as a Senior Software Engineer at Buraq Technologies, Wah Cantt, Pakistan. He received his BS CS degree from PMAS-Arid Agriculture University, Rawalpindi, Pakistan.